\documentclass[
twocolumn,
]{ceurart}

\usepackage{listings}
\usepackage{setspace}
\usepackage{booktabs}
\usepackage{caption}
\usepackage{subcaption}
\usepackage{multirow}
\begin{document}

\copyrightyear{2022}
\copyrightclause{Copyright for this paper by its authors.
  Use permitted under Creative Commons License Attribution 4.0
  International (CC BY 4.0).}

\conference{IJCAI 2023 Workshop on Deepfake Audio Detection and Analysis (DADA 2023), August 19, 2023, Macao, S.A.R}

\title{Adaptive Fake Audio Detection with Low-Rank Model Squeezing}


\author[1,2]{Xiaohui Zhang}[
orcid=0000-0002-9949-5415,
email=21120320@bjtu.edu.cn]
\address[1]{State Key Laboratory of Multimodal Artificial Intelligence System, Institute of Automation, Chinese Academy of Sciences}
\address[2]{School of Computer and Information Technology, Beijing Jiaotong University}

\author[1]{Jiangyan Yi}[email=jiangyan.yi@nlpr.ia.ac.cn]
\cormark[1]
\author[4]{Jianhua Tao}[email=jhtao@tsinghua.edu.cn]
\cormark[1]
\author[1,3]{Chenglong Wang}[email=chenglong.wang@nlpr.ia.ac.cn]
\address[3]{University of Science and Technology of China}
\author[1,5]{Le Xu}[email=lexu@nlpr.ia.ac.cn]
\author[1]{Ruibo Fu}[email=ruibo.fu@nlpr.ia.ac.cn]
\address[4]{Department of Automation, Tsinghua University}
\address[5]{School of Artificial Intelligence, University of Chinese Academy of Sciences}

\cortext[1]{Corresponding author.}

\begin{abstract}
The rapid advancement of spoofing algorithms necessitates the development of robust detection methods capable of accurately identifying emerging fake audio. Traditional approaches, such as finetuning on new datasets containing these novel spoofing algorithms, are computationally intensive and pose a risk of impairing the acquired knowledge of known fake audio types. To address these challenges, this paper proposes an innovative approach that mitigates the limitations associated with finetuning. We introduce the concept of training low-rank adaptation matrices tailored specifically to the newly emerging fake audio types. During the inference stage, these adaptation matrices are combined with the existing model to generate the final prediction output.
Extensive experimentation is conducted to evaluate the efficacy of the proposed method. The results demonstrate that our approach effectively preserves the prediction accuracy of the existing model for known fake audio types. Furthermore, our approach offers several advantages, including reduced storage memory requirements and lower equal error rates compared to conventional finetuning methods, particularly on specific spoofing algorithms.
\end{abstract}


\maketitle

\section{Introduction}
In recent years, there has been a significant concern surrounding the issue of audio forgery attacks. Detection models for detecting fake audio, based on handcrafted features \cite{CQCC, ASGCAS} and large-scale pre-trained models \cite{Wangxipretrain}, have achieved promising performance on multiple competition datasets \cite{ASVspoof2015, ASVspoof2017,ASVspoof2019,ASVspoof2021, ADD2022}. However, when faced with audio generated by spoofing algorithms that were not encountered during training, these models experience a significant decrease in their discrimination accuracy \cite{DFWF,Empirical21}. This issue has become one of the crucial factors hindering the practical application of fake audio detection models. As new audio spoofing techniques continue to emerge, there is a need for a method to improve the discriminative ability of fake audio detection models against new spoofing attacks.\par
The most intensive way to improve the detection accuracy of the model against new spoofing algorithms is to finetune the model on a new dataset including those unseen types of fake audio. However, finetuning the model on the new dataset can disrupt the knowledge model learned from the old dataset, leading to a decrease in the recognition accuracy of the model for fake audio generated by known spoofing algorithms, which is known as catastrophic forgetting \cite{EWC,DFWF}. In addition, if the model has a large number of parameters, 
simultaneously fine-tuning will not only require a high training time and computational memory consumption, but also result in a large saved model that is difficult to use in scenarios with storage space limitations. \par
To mitigate the detrimental impact of fine-tuning on acquired knowledge, we propose a novel training approach based on Low-Rank Adaption (LoRA) \cite{LoRA}. Our method tackles the issue of poor performance of the model on unseen types of fake audio. The core of our approach lies in training two low-rank adaptive matrices rather than finetuning the whole model for improving the recognized accuracy of the unseen fake audio. 
During training on the new dataset that includes those unseen fake audio, we load the source model (SoM), which is a saved model training on the old dataset, and freeze all its parameters. This allows us to solely focus on training two adaptive matrices, namely $\mathbf{A}$ and $\mathbf{B}$, as introduced by the LoRA algorithm. When performing inferences on the new dataset, we load the SoM together with the two adaptive matrices. Conversely, when dealing with the old dataset, we only load the SoM. Compared to fine-tuning, our method abstains from altering the parameters of the SoM, effectively evading the risk of damage to the knowledge obtained from known instances of fake audio in the old dataset. Additionally, our approach boasts an advantage in terms of storage memory consumption, as it only necessitates the storage of the two low-rank adaptive matrices $\mathbf{A}$ and $\mathbf{B}$ for the new and unseen fake audio. Furthermore, experimental results demonstrate that our method achieves lower equal error rates (EER) \cite{ASVspoof2015} on certain types of unseen fake audio compared to finetuning.\par
\textbf{Contribution}:
We propose a method based on Low-Rank Adaption to address the issue of low recognized accuracy when models encounter new and unknown types of fake audio in fake audio detection. Compared to the commonly used fine-tuning approach, our method requires lower storage space and avoids forgetting the knowledge learned from existing known types of fake audio. Additionally, experimental results demonstrate that our method achieves higher recognized accuracy for certain unknown types of spoofing algorithms compared to finetuning.
\section{Related Work}
\begin{figure*}[t]
\begin{center}
\begin{subfigure}[ht]{0.49\linewidth}
\begin{center}
\includegraphics[width=1.\linewidth]{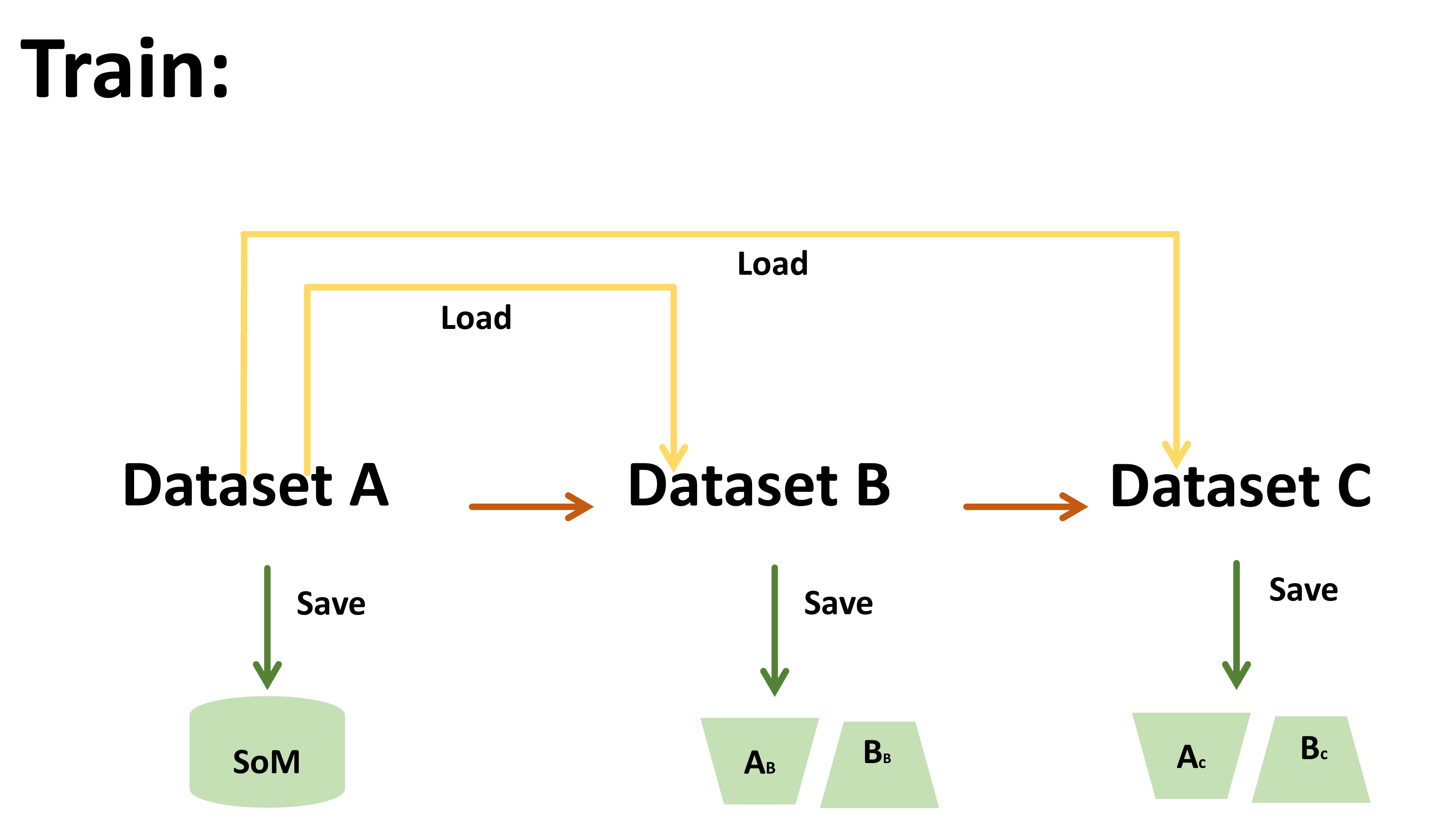}
\caption{}
\label{Atrain}
\end{center}
\end{subfigure}
\begin{subfigure}[ht]{0.49\linewidth}
\begin{center}
\includegraphics[width=1.\linewidth]{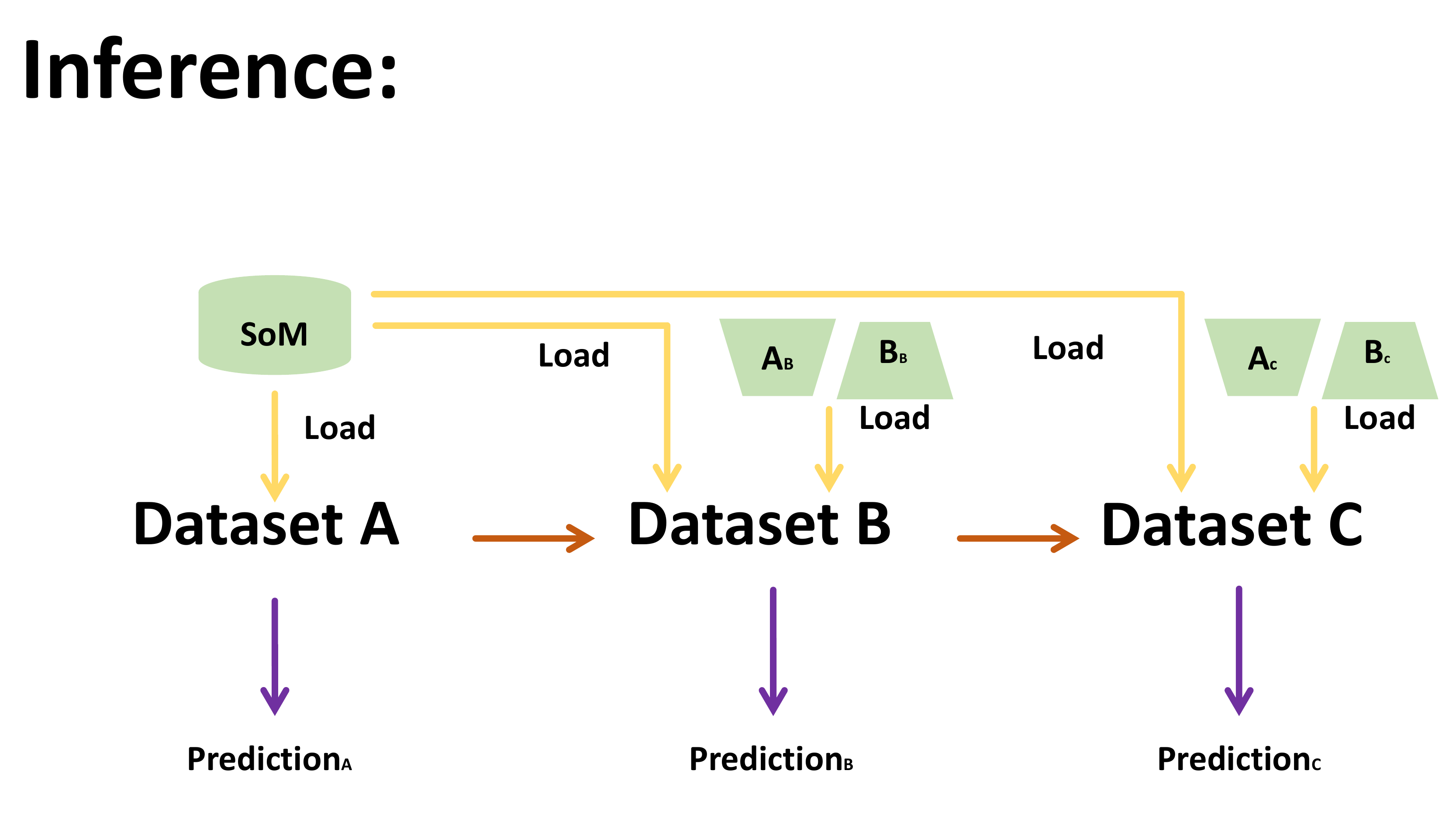} 
\caption{}
\label{Ainfer}
\end{center}
\end{subfigure}
\caption{The training (a) and inference (b) process of our method. Dataset A represents the dataset consisting of known types of fake audio currently. Two additional datasets B and C contain new types of fake audio that are not presented in dataset A. They can be viewed as two datasets we collect after a certain time period when the training process has completed on known types of fake audio and some new and unseen spoofing algorithms have been proposed.}
\label{Art}
\end{center}
\end{figure*}
Low-Rank Adaption (LoRA)\cite{LoRA} is a method proposed to significantly reduce GPU and storage memory consumption when finetuning large-scale pre-trained transformer models for specific tasks. The core idea of LoRA is that the learned over-parametrized models actually reside on a low intrinsic dimension. Therefore, for each specific downstream task, LoRA introduces two low-rank matrices, $\mathbf{A}$ and $\mathbf{B}$, to replace the entire model during training. In each downstream task, LoRA first loads the large-scale pre-trained model but freezes all its parameters. Then, it initializes matrices $\mathbf{A}$ and $\mathbf{B}$ as zero matrices and trains only these two matrices using the training dataset of the downstream task. During model inference, LoRA simultaneously loads the large-scale pre-trained model and the two matrices $\mathbf{A}$ and $\mathbf{B}$ trained specifically for that downstream task. Compared to existing methods that introduce adapter layers \cite{al-1,al-2,al-3,al-4}, LoRA exhibits lower inference latency \cite{LoRA}. This method also reduced the number of trainable parameters in the training of the large-scale language model GPT-3 \cite{GPT-3} by a factor of 10,000 and decreased GPU memory requirements by 3 times.
\section{Methodology}
When facing unknown types of fake audio generated by unknown algorithms, the accuracy of deep neural networks would significantly decrease compared to the known types included in the training set. The finetuning-based methods may damage the learned knowledge of the model, leading to a reduction in the detection accuracy of known types of fake audio. To address this problem, we propose a new method based on LoRA to improve the recognized ability of the model to detect unknown types of fake audio. The training and inference processes of our method are illustrated in Fig \ref{Atrain} and \ref{Ainfer}, respectively. We consider a model designed for fake speech detection, where there exists an initial dataset A containing some fake audio generated by known spoofing algorithms, and two additional datasets B and C, both of which contain new types of fake audio not present in dataset A. We first train a source model (SoM) on dataset A. As SoM has not seen the new types of fake audio in datasets B and C, it can achieve good recognition performance on dataset A, but its performance would significantly decrease on datasets B and C. To improve the detection accuracy of new types of fake audio in dataset B, we load the saved SoM trained on dataset A, freeze all its parameters, and introduce two low-rank adaptive matrices $\mathbf{A}_B$ and $\mathbf{B}_B$ specifically designed for dataset B. Both $\mathbf{A}_B$ and $\mathbf{B}_B$ are initialized as all-zero matrices with rank $r_B$, which is much lower than the rank of SoM. During the training process on dataset B, we simultaneously feed data into both SoM and the adaptive matrices $\mathbf{A}_B$ and $\mathbf{B}_B$. The outputs from both components are summed to generate the output of the model $h_{model}$, as shown in Equation \ref{Train_som-ab}. 
\begin{equation}
    h_{model} = \mathbf{W}_{SoM}x + \mathbf{A_B B_B}x
\label{Train_som-ab}
\end{equation}
where the $x$ represents the input batch of data and the $h_{model}$ is the output state of the model. While the parameters of SoM remain unchanged during the training on dataset B, we optimize the parameters of the two low-rank adaptive matrices $\mathbf{A}_B$ and $\mathbf{B}_B$ to learn new features for the new types of fake audio and optimize the detection performance of the model on these types. Once the training on dataset B is completed, we only need to save the two low-rank adaptive matrices, $\mathbf{A}_B$ and $\mathbf{B}_B$, instead of the entire model. The same process is repeated on dataset C, allowing us to train two additional low-rank adaptive matrices, $\mathbf{A}_C$ and $\mathbf{B}_C$, specifically tailored to the new types of fake audio in dataset C.\par
Our method follows a similar inference process across different datasets, as shown in Fig \ref{Ainfer}. When the model predicts a fake audio type belonging to Dataset A, we only load the SoM. Since the parameters of the SoM are frozen and not involved in the training process on Datasets B and C, the detection performance on Dataset A is not disrupted by the learned features from the new fake audio in Datasets B and C. When the model predicts a fake audio type from Dataset B, both the source model SoM and the low-rank adaptive matrices $\mathbf{A}_B$ and $\mathbf{B}_B$ are loaded into the model, and the output follows Equation \ref{Train_som-ab}. Although the parameters of the SoM are not updated during training, the model can learn the features of the new fake audio type by training the parameters of the two adaptive matrices. As a result, compared to using only the source model SoM, the detection accuracy of the model is significantly improved. Similarly, when the model faces fake audio types from Dataset C, we can load the SoM and the adaptive matrices $\mathbf{A}_C$ and $\mathbf{B}_C$ trained specifically for Dataset C.\par
Overall, our algorithm provides a low-cost incremental learning method for the model. As audio spoofing algorithms continue to evolve, we can view Dataset A as consisting of the known spoofing algorithms and set a time period $t_{ud}$, after which we collect new fake audio generated by emerging spoofing algorithms to build Dataset B. We apply our method to incrementally learn the model on Dataset B. After another time period of $2t_{ud}$, we repeat the process to construct Dataset C and perform incremental learning on Dataset C. Through this approach, we enable our model to achieve self-incremental learning within limited storage space, thus defending against emerging attacks from new spoofing algorithms.

\section{Experiment}
\subsection{Datasets}
Three fake audio datasets are selected for our experiments, including the ASVspoof2019LA \cite{ASVspoof2019}, ASVspoof2015 \cite{ASVspoof2015}, and In-the-Wild \cite{ITW}.
All of the experiments are trained on training sets and evaluated on evaluation sets in these datasets.\par
\textbf{ASVspoof2019LA} is a dataset widely used in the field of fake audio detection. It was created as part of an international challenge that aimed to evaluate the performance of automatic speaker verification systems in detecting spoofing attacks. The dataset consists of a large collection of both genuine and spoofed speech recordings, where spoofed speech refers to artificially generated or manipulated audio designed to deceive speaker verification systems. \par
\textbf{ASVspoof2015} is another important dataset used in fake audio detection research. The dataset contains both genuine and spoofed speech recordings, with various types of spoofing attacks, such as speech synthesis, voice conversion, and replay attacks. ASVspoof2015 offers a diverse range of spoofing techniques, making it a valuable resource for studying and developing robust countermeasures against fake audio.\par
\textbf{In-the-Wild} is a commonly used collection of real-world audio recordings that encompass a broad range of environments and scenarios. Unlike the aforementioned datasets that focus on specific spoofing attacks, In-the-Wild captures audio data from various sources and situations encountered in everyday life. This dataset aims to simulate the challenges faced by fake audio detection systems when dealing with uncontrolled and unpredictable acoustic conditions. We divide the genuine and fake audios of the In-the-Wild dataset into two subsets. One-third is used to build the training set, and the rest is used as the evaluation set.
\subsection{Experimental Setup}
\label{ES}
In our experiments, the Low-Level Cepstral Coefficients (LFCC) \cite{LFCC} feature has been selected as the input feature extracted from each audio. The classifier is the Squeeze-and-Excitation Network (SENet) \cite{SENet} with three sub-layers. All of them include three basic blocks introduced by the SENet. There is one conv2d layer before each sub-layer. The input dim and output dim of the first conv2d are 1 and 128, respectively. The second conv2d and the third have input and output dim 128 and 256, 256 and 512, respectively. The kernel sizes of them are 9, 7, and 5. The batch size is 64 and the optimizer is Adam optimizer with a learning rate of 0.001.
\subsection{Only trained on ASVspoof2019LA}
We first test the recognized performance of the deep neural network against fake audio generated by known and unknown algorithms, respectively. We consider the datasets ASVspoof2019LA, ASVspoof2015, and In-the-Wild as datasets A, B, and C, respectively, as shown in Fig \ref{Art}. We train the model only on the training set of ASVspoof2019LA and evaluate it on the evaluation sets of the three datasets. The experimental results are shown in Table \ref{train2019}. The results indicate that the model has high accuracy when faced with known types of fake audio that have appeared in the training set, but its recognized performance will degrade considerably when faced with fake audio generated by new and unknown spoofing algorithms.
\begin{table}[t]
\begin{center}
\caption{The EER(\%) on the evaluation set of each datasets. The model SoM is only trained on the ASVspoof2019LA dataset.}
\label{train2019}
\resizebox{!}{0.47cm}{
\begin{tabular}{cccc}
\toprule[1.pt]
\multicolumn{1}{c}{\bf Dataset}                &
\multicolumn{1}{c}{$\mathbf{ASVspoof2019LA}$}                   &
\multicolumn{1}{c}{$\mathbf{ASVspoof2015}$} &
\multicolumn{1}{c}{\textbf{In-the-Wild}} 
\\ 
\midrule[0.5pt]
EER(\%)        &   $6.51$   &   $51.77$ & $51.71$        \\
\bottomrule[1.pt]
\end{tabular}
}
\end{center}
\end{table}
\begin{table}
\caption{The comparison of the recognized performance between finetuning and our method. (a) and (b) are the evaluation EER (\%) of the SoM after training on the ASVspoof2015 and In-the-Wild, respectively.}
\begin{center}
\begin{subtable}[t]{0.98\linewidth}
\begin{center}
\caption{}
\label{two-1}
\resizebox{!}{0.8cm}{
\begin{tabular}{ccc}
\toprule[1.pt]
\multicolumn{1}{c}{\bf EER(\%)}  &   \multicolumn{1}{c}{$\mathbf{ASVspoof2019LA}$}  &    \multicolumn{1}{c}{$\mathbf{ASVspoof2015}$}
\\ 
\midrule[0.5pt]
SoM        &   $6.51$        &   $51.77$            \\
\midrule[0.5pt]
Finetuning &   $49.03$       &   $5.06$             \\
\textbf{Our method} &   $\mathbf{6.51}$        &   $\mathbf{2.38}$             \\
\bottomrule[1.pt]
\end{tabular}}
\end{center}
\end{subtable}
\begin{subtable}[t]{0.98\linewidth}
\begin{center}
\caption{}
\label{two-2}
\resizebox{!}{0.8cm}{
\begin{tabular}{ccc}
\toprule[1.pt]
\multicolumn{1}{c}{\bf EER(\%)}  &   \multicolumn{1}{c}{$\mathbf{ASVspoof2019LA}$}  &    \multicolumn{1}{c}{\textbf{In-the-Wild}}
\\ 
\midrule[0.5pt]
SoM        &   $6.51$        &   $51.71$            \\
\midrule[0.5pt]
Finetuning &   $33.05$       &   $0.75$             \\
\textbf{Our method} &   $\mathbf{6.51}$        &   $\mathbf{1.25}$             \\
\bottomrule[1.pt]
\end{tabular}}
\end{center}
\end{subtable}
\end{center}
\label{cm-2}
\end{table}
\begin{table}[t]
\begin{center}
\caption{The total parameters count of our method and finetuning in the training process.}
\label{SM}
\resizebox{!}{0.52cm}{
\begin{tabular}{cccc}
\toprule[1.pt]
\multicolumn{1}{c}{\bf Storage Memory (M)}                &
\multicolumn{1}{c}{$\mathbf{ASVspoof2019LA}$}                   &
\multicolumn{1}{c}{$\mathbf{ASVspoof2015}$} &
\multicolumn{1}{c}{\textbf{In-the-Wild}} 
\\ 
\midrule[0.5pt]
Finetuning     &   $23.61$   &   $23.61$    & $23.61$        \\
\textbf{Our method}     &   $23.61$   &   $\mathbf{2.44}$ & $\mathbf{2.44}$        \\
\bottomrule[1.pt]
\end{tabular}}
\end{center}
\end{table}
\begin{table}[t]
\begin{center}
\caption{The comparison of the recognized performance between finetuning and our method. The evaluation EER (\%) in the following table is the SoM after first training on the ASVspoof2015 and then training on the In-the-Wild.}
\label{cm-3}
\resizebox{!}{0.75cm}{
\begin{tabular}{cccc}
\toprule[1.pt]
\multicolumn{1}{c}{\bf EER(\%)}                &
\multicolumn{1}{c}{$\mathbf{ASVspoof2019LA}$}  &
\multicolumn{1}{c}{$\mathbf{ASVspoof2015}$}    &
\multicolumn{1}{c}{\textbf{In-the-Wild}}     \\ 
\midrule[0.5pt]
SoM &   $6.51$  &   $51.77$     &   $51.71$ \\
\midrule[0.5pt]
Finetuning    &   $35.31$   &   $45.13$   & $1.39$  \\
\textbf{Our method}    &   $\mathbf{6.51}$   &   $\mathbf{2.38}$   & $\mathbf{1.25}$  \\
\bottomrule[1.pt]
\end{tabular}}
\end{center}
\end{table}
\subsection{The comparison on EER between finetuning and our method on learning between two datasets}
In this section, we compare the recognized performance between our method and finetuning on two-dataset learning condition. We set two experimental situations: the first is the model first trained on the ASVspoof2019LA and then trained on the ASVspoof2015; the second is the model first trained on the ASVspoof2019LA and then trained on the In-the-Wild. The results of these two experiments are shown in Table \ref{two-1} and \ref{two-2}, respectively. From the second column of the two tables, we can easily observe that training on the new dataset is really beneficial for the detection of new fake audio generated from new spoofing algorithms. However, from the comparison in the first column, we can observe that finetuning on the new dataset will definitely disrupt the learned knowledge from the known types of fake audio ($6.51 \rightarrow 49.03, 6.51 \rightarrow 33.05$). Compared to finetuning, our method freeze the parameters of the SoM and only trained two adaptive matrices to learn new knowledge from the new dataset. In this case, the recognized performance of the known fake audio types will remain unchanged even after training on the new dataset \ref{Ainfer}, which is evaluated in our results shown in Table \ref{cm-2}. From the comparison on the learning performance between our method and finetuning in the second column, we can also see that our method achieves a higher recognized accuracy in ASVspoof2019LA $\rightarrow$ ASVspoof2015, which shows that our method has a positive effect on the learning on specific unknown spoofing algorithms.
\subsection{The comparison on EER between finetuning and our method on learning among three datasets}
To evaluate the effectiveness of our method in multi-dataset learning, we also compare the recognized performance between our method and finetuning on three datasets learning condition. In our experiment, we first trained our model in the ASVspoof2019LA and saved the completed source model SoM. After that, we trained the SoM first on the ASVspoof2015 and then on the In-the-Wild, and saved the adaptive matrices $\mathbf{A_B}, \mathbf{B_B}$ and $\mathbf{A_C}, \mathbf{B_C}$, respectively. The inference process is shown in Fig \ref{Ainfer} and the comparison result is illustrated in Table \ref{cm-3}. From the comparison of the first two rows in the result, we can observe that finetuning on new datasets will reduce the recognized accuracy on old datasets, which shows that the detection accuracy of the known types of fake audio will considerably decrease after finetuning on the unknown types of fake audio. However, we can see that our method still remains unchanged in old datasets ASVspoof2019LA and ASVspoof2015 and achieves lower EER than finetuning on the final dataset.

\subsection{The comparison on storage memory between finetuning and our method}
In order to improve the recognition performance of the model, we train it according to the process illustrated in Fig \ref{Art}. After training on the new datasets ASVspoof2015 and In-the-Wild, we compare the storage memory between the whole model and adaptive matrices saved by finetuning and our method, respectively, which has been shown in Table \ref{SM}. The experimental result shows that our method achieves a marked success in squeezing storage memory. Under the setting illustrated in Sec \ref{ES}, our method greatly reduces the number of trainable parameters and the storage memory requirement by about 30 times, which makes the model can be easily applied in many strict memory constraint situations.

\section{Conclusion}
In this paper, we propose a method to address the problem of low detection accuracy of models facing newly emerging fake audio generated by new spoofing algorithms. In the training process, we 
train two low-rank adaptation matrices $\mathbf{A}$ and $\mathbf{B}$ specifically for these new types of fake audio. During inference, we simultaneously load the existing model and these adaptation matrices, and combine their prediction outputs as our final prediction output. The experimental results demonstrate that our method does not degrade the prediction accuracy of the existing model for known types of fake audio because the existing model parameters are not modified during training on the new dataset. Moreover, our method has a lower storage memory requirement and lower equal error rates on some specific spoofing algorithms compared to finetuning. These findings encourage further investigation into countering the ever-evolving landscape of audio spoofing while maintaining the learned knowledge of known types of fake audio.
\begin{acknowledgments}
This work is supported by the National Natural Science Foundation of China (NSFC) (No.61831022, No.U21B2010, No.62101553, No.61971419, No.62006223, No.62276259, No.62201572, No. 62206278), Beijing Municipal Science and Technology Commission, Administrative Commission of Zhongguancun Science Park No.Z211100004821013, Open Research Projects of Zhejiang Lab (NO. 2021KH0AB06).
\end{acknowledgments}

\bibliography{sample-2col}

\end{document}